\documentclass[%
pra%
 ,twocolumn%
 ,secnumarabic%
,amssymb, amsmath,nobibnotes, aps, pra]{revtex4}
\usepackage{docs}%
\usepackage{graphicx}% added 121030
\usepackage{bm}%
\expandafter\ifx\csname package@font\endcsname\relax\else
 \expandafter\expandafter
 \expandafter\usepackage
 \expandafter\expandafter
 \expandafter{\csname package@font\endcsname}%
\fi

\begin{document}

\title{Quantum noise reduction techniques in KAGRA}
\author{K. Somiya}
\email{somiya@phys.titech.ac.jp}
\affiliation{Department of Physics, Tokyo Institute of Technology, 2-12-1 Oh-okayama Meguro Tokyo 152-8551}
\date{Jul 2019}

\begin{abstract}
KAGRA is the first large-scale gravitational-wave detector with cryogenic test masses. Its target sensitivity is limited mostly by quantum noise in the observation frequency band owing to the remarkable reduction of thermal noise at cryogenic temperatures. It is thus essential to reduce quantum noise, and KAGRA is designed to implement two quantum noise reduction techniques. KAGRA has already started considering an upgrade plan, in which a few more new quantum noise reduction techniques will be incorporated. In this article, we report the currently implemented quantum noise reduction techniques for KAGRA and those that will be implemented in the near future.
\end{abstract}

\pacs{0.00}

\maketitle

\section{Introduction}

A gravitational-wave detector is a large-scale laser interferometer that can probe a displacement as small as $\sim\!10^{-21}$\,m. To achieve such a high sensitivity, the interferometer has to be carefully designed. There are three fundamental noise sources that limit the sensitivity of the currently operating detectors. The first noise source is seismic and seismic Newtonian noise, which can limit the sensitivity at very low frequencies ($f<\sim\!10\,$Hz). Another noise source is thermal noise of the mirrors and the suspensions, which can limit the sensitivity at around 100\,Hz. The third noise source is quantum noise, which consists of quantum radiation pressure noise, limiting the sensitivity at low frequencies ($10\,\mathrm{Hz}<f<100\,\mathrm{Hz}$), and photon shot noise, limiting the sensitivity at high frequencies (above a few hundred hertz). For KAGRA, the large-scale cryogenic underground detector in Japan~\cite{KAGRA}, mirror thermal noise is lower, but quantum noise is higher than the other gravitational-wave detectors used in the world. Quantum radiation pressure noise in KAGRA is high because the sapphire mirror cannot be made as large as the fused silica glass mirror used in the other detectors. Photon shot noise in KAGRA is high because the laser power in KAGRA cannot be increased to the high values that are possible in other detectors if the mirror temperature is to be maintained in the range $20\textendash23$\,K.

In fact, there are ways to decrease quantum noise without increasing the mass or the laser power. One method is to implement the quantum non-demolition technique. The key idea is to change the readout quadrature at the photo detection so that a part of the amplitude quadrature fluctuation is coherently subtracted, and quantum radiation pressure is reduced in the vicinity of a certain frequency~\cite{BAE}. Another method is to implement the optical spring. The key idea is to detune the signal-recycling cavity (SRC) length so that a phase signal is partly converted to an amplitude signal to drive the mirror via radiation pressure. This restoring force creates a kilometer-scale spring between the two distant test masses~\cite{OpticalSpring}. The two techniques can be combined; the optimization of the readout quadrature would not then aim to coherently subtract radiation pressure noise but would aim to select a good frequency response of the optical spring for the anticipated gravitational-wave sources. These techniques are to be implemented in the current KAGRA.

Although a concrete plan to further upgrade KAGRA is not yet fixed, it is clear that heavier sapphire test masses with low photon absorption are required. An alternative way to effectively increase the test mass and lower quantum radiation pressure noise is to use a frequency-dependent squeezing technique~\cite{Kimble}. The key idea is to inject a frequency-independent squeezed vacuum into a long and high-finesse, so-called {\it filter cavity} so that the squeezing angle can rotate at a frequency where the quantum radiation pressure noise level equals the photon shot noise level. With frequency-dependent squeezing, both quantum radiation pressure noise and photon shot noise decrease as though both the mirror mass and the laser power have increased. It is also an option that can be used to simply increase the laser power while sacrificing the cryogenic temperature and/or using thicker suspension wires. With the current suspension wire, which is 1.6\,mm in diameter and 35\,cm long, and with the mirror temperature as high as 29\,K, the laser power at the beam splitter can be as high as 3.4\,kW. With a new suspension wire that is 2.5\,mm in diameter and 20\,cm long, the mirror temperature can be as low as 20\,K, and the laser power at the beam splitter can be maintained at 3.4\,kW. 
%Since it will take long to develop a large sapphire mirror while it is better to implement some advanced techniques by the time other detectors complete upgrades and start running as 2.5-th generation detectors, KAGRA is planning to conduct the first phase of the upgrade plan, or called {\it KAGRA+}. The use of the frequency-independent squeezing and the higher power laser will be a good strategy. 
Another idea is to extend the signal-recycling cavity length to a few hundred meters~\cite{longSRC}. The optical resonance of the signal is moved to a few kilohertz, which is a good frequency to observe a gravitational wave from the merger of two neutron stars. The shot noise level decreases in the broadband with frequency independent squeezing. This can be a promising technique if a space can be found in which to install the few-hundred-meters-long cavity. Even with the current SRC length, the long SRC effect can be achieved if we increase the finesse of the arm cavity and the signal-recycling gain.

The structure of this article is as follows. In Sec.~\ref{sec:2}, we briefly explain quantum noise by using a two-photon formalism and input-output relations. In Sec.~\ref{sec:3}, we describe the quantum noise reduction techniques that are to be implemented in KAGRA. In Sec.~\ref{sec:4}, we introduce a few advanced quantum noise reduction techniques that are to be incorporated in the upgrade plan of KAGRA for the near future~\cite{WP}.

\section{Quantum noise}\label{sec:2}

\begin{figure}[htbp]
	\begin{center}
		\includegraphics[width=8cm]{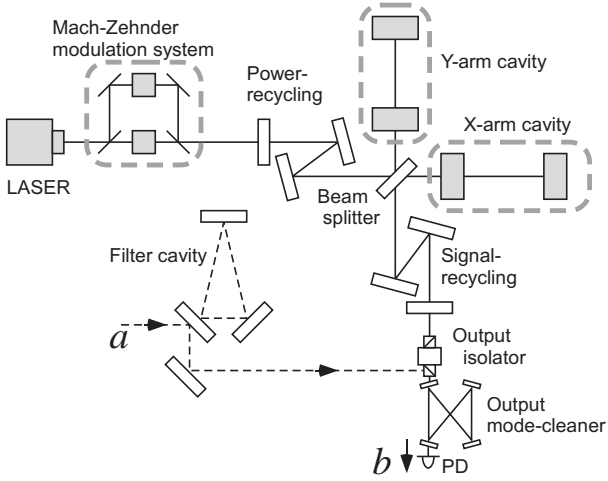}
	\caption{Schematic of the KAGRA interferometer. The quantum noise comes from the vacuum field ``$\it a$'' entering from the anti-symmetric port of the interferometer through the output isolator. The signal is included in the field ``$\it b$'' coming out through the output mode-cleaner. The Mach-Zehnder modulation system is used to control the detuned signal-recycling cavity (SRC) (Sec.~\ref{sec:3}), and the filter cavity is used to realize frequency-dependent squeezing (Sec.~\ref{sec:4}).}
	\label{fig:IFO}
	\end{center}
\end{figure}

Figure~\ref{fig:IFO} shows a schematic of KAGRA. Some elements not mentioned in this article have been omitted from the schematic. The detector is based on the Michelson interferometer with an optical cavity in each arm and a couple of folded recycling cavities. The filter cavity, output mode-cleaner, output isolator, and Mach{\textendash}Zehnder modulation system are also depicted for later explanations.

Let $\omega_0$ denote the angular frequency of the carrier light. The electromagnetic field in the Heisenberg picture is given by
\begin{eqnarray}
E(t)=\sqrt{\frac{2\pi\hbar\omega_0}{{\cal A}c}}e^{-i\omega_0t}\!\!\int_0^\infty\!\!\!\!\left[a_+e^{-i\Omega t}+a_-e^{i\Omega t}\right]\frac{d\Omega}{2\pi}+\mathrm{H.C.},\nonumber\\
\end{eqnarray}
with $a_\pm$ being the annihilation operator at $\omega_0\pm\Omega$. Here, $\hbar$ is the Planck constant, $c$ is the speed of light, ${\cal A}$ is the cross-sectional area of the beam, and H.C. denotes the Hermitian conjugate. The amplitude quadrature with subscript ``1'' changes in-phase, and the phase quadrature with subscript ``2'' changes with a $\pi/2$ phase delay with respect to the carrier light: 
\begin{eqnarray}
a_1=\frac{a_++a_-}{\sqrt{2}}\,,\ \ a_2=\frac{a_+-a_-}{\sqrt{2}i}\,.
\end{eqnarray}
These fields satisfy the following commutation relations: 
\begin{eqnarray}
[a_1(\Omega),\,a_2^\dagger(\Omega')]=[a_2(\Omega),\,a_1^\dagger(\Omega')]=2\pi i\delta(\Omega-\Omega')\,. 
\end{eqnarray}
Denote the electromagnetic field entering the interferometer from the anti-symmetric port as $(a_1,\,a_2)$ and the field emerging from the interferometer at the anti-symmetric port as $(b_1,\,b_2)$; after some algebraic manipulation we obtain~\cite{OpticalSpring}
\begin{eqnarray}
\left (
\begin{array}{@{\,}c@{\,}}
b_1\\
b_2
\end{array}\!\right )=
\frac{1}{M}\left[\left (\!
\begin{array}{@{\,}cc@{\,}}
A_{11}&A_{12}\\
A_{21}&A_{22}
\end{array}\!\right )\!\!
\left (
\begin{array}{@{\,}c@{\,}}
a_1\\
a_2
\end{array}\!\right )e^{2i(\beta+\Phi)}\right.\nonumber\\
\ \ \ \left.
+\sqrt{2{\cal K}}t_s\frac{h}{h_\mathrm{SQL}}\left (\!
\begin{array}{@{\,}c@{\,}}
D_1\\
D_2
\end{array}\!\right )e^{i(\beta+\Phi)}\right]\nonumber\\
\end{eqnarray}
with
\begin{eqnarray}
M&=&1-2r_s\left(\cos{2\phi}+\frac{\cal K}{2}\sin{2\phi}\right)e^{2i(\beta+\Phi)}+r_s^2e^{4i(\beta+\Phi)},\nonumber\\
A_{11}&=&A_{22}\nonumber\\
&=&(1+r_s^2)\left(\cos{2\phi}+\frac{\cal K}{2}\sin{2\phi}\right)-2r_s\cos{[2(\beta+\Phi)]},\nonumber\\
A_{12}&=&-t_s^2(\sin{2\phi}+{\cal K}\sin^2{\phi}),\nonumber\\
A_{21}&=&t_s^2(\sin{2\phi}-{\cal K}\cos^2{\phi}),\label{eq:inputoutput}\\
D_1&=&(1+r_se^{2i(\beta+\Phi)})\sin{\phi},\nonumber\\
D_2&=&(1-r_se^{2i(\beta+\Phi)})\cos{\phi}.\nonumber
\end{eqnarray}
Here, $r_s$ and $t_s$ are the amplitude reflectivity and transmittance of the signal-recycling mirror, $\phi$ is the phase rotation gained by the carrier in the SRC ($\phi=[\omega_0\ell/c]_\mathrm{mod 2\pi}$), $\Phi$ is the phase rotation gained by a signal sideband in the SRC ($\phi=[\Omega\ell/c]_\mathrm{mod 2\pi}$), and $h$ is the gravitational-wave signal in strain. In addition, ${\cal K}$ is the optomechanical coupling coefficient of the Fabry-Perot Michelson interferometer, $\beta$ is the phase rotation gained by the signal sideband in the arm cavity, and $h_\mathrm{SQL}$ is the standard quantum limit (SQL) of the strain measurement of the interferometer:
\begin{eqnarray}
{\cal K}=\frac{8\omega_0I_0}{mL^2\Omega^2(\gamma^2+\Omega^2)},\ \tan{\beta}=\frac{\Omega}{\gamma},\ h_\mathrm{SQL}=\sqrt{\frac{8\hbar}{m\Omega^2L^2}}\nonumber\\
\end{eqnarray}
where $I_0$ is the laser power at the beamsplitter, $L$ is the arm cavity length, $m$ is the mass of the mirror, and $\gamma=Tc/4L$ is the cavity pole angular frequency with $T$ being the transmittance of the input mirror.

Let us first calculate the quantum noise level for a simple case without the signal-recycling mirror. With $r_s=0$, $t_s=1$, $\phi=0$, and $\Phi=0$, the noise term in the $b_2$ component is $(-{\cal K}a_1+a_2)e^{2i\beta}$, and the signal term is $\sqrt{2\cal K}e^{i\beta}h/h_\mathrm{SQL}$. Taking into account the fact that the single-sided spectral density and cross-spectral density are given by~\cite{Kimble}
\begin{eqnarray}
S_{a_1}(f)=S_{a_2}(f)=1,\ \ \ S_{a_1a_2}(f)=0\,,\label{eq:coherent}
\end{eqnarray} 
we obtain the noise spectral density of the interferometer as
\begin{eqnarray}
S_h(f)=\frac{h_\mathrm{SQL}^2}{2}\left(\frac{1}{\cal K}+{\cal K}\right)\,.\label{eq:Sh}
\end{eqnarray} 
On the right-hand side of Eq.~(\ref{eq:Sh}), the first term in brackets represents the shot noise level, which decreases with the laser power $I_0$, and the second term in brackets represents the radiation pressure noise, which increases with the laser power $I_0$. As a result, there exists a lower limit of the noise spectral density that cannot be exceeded simply by changing the laser power: 
\begin{eqnarray}
S_h(f)\geq h_\mathrm{SQL}^2\,.
\end{eqnarray} 
The limit is called the SQL~\cite{SQL}.

\begin{figure}[htbp]
	\begin{center}
		\includegraphics[width=8cm]{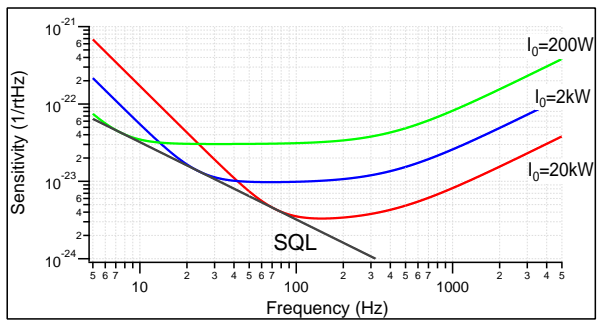}
	\caption{Quantum noise levels of a power-recycled Fabry-Perot Michelson interferometer with different power levels at the beamsplitter. The noise levels do not exceed the standard quantum limit (SQL) (dashed line).}
	\label{fig:SQL}
	\end{center}
\end{figure}

Figure~\ref{fig:SQL} shows three example quantum noise spectra with different $I_0$ values . The transmissivity of the input mirror is set $T=0.1$. The other parameters are not changed from the ones for KAGRA: $m=23$~kg and $L=3$~km. The cavity pole is calculated to be $\gamma=Tc/4L\simeq2\pi\times400$~Hz. The quantum noise spectrum reaches the SQL at a certain frequency when ${\cal K}$ reaches unity. The cavity pole should preferably be set higher than the SQL touching frequency.

Next, let us introduce the signal recycling mirror. We shall set the phase rotations $\phi$ to $\pi/2$; which is the broadband resonant sideband extraction (RSE) configuration~\cite{Mizuno}. The phase rotation $\Phi$ is still negligible. The noise term and the signal term in the $b_2$ component contain $A_{21}/M$, $A_{22}/M$, and $t_sD_2/M$, which are given as
\begin{eqnarray}
\frac{A_{21}}{M}&=&-\frac{t_s^2}{(1+r_se^{2i\beta})^2}{\cal K},\nonumber\\
\frac{A_{22}}{M}&=&\frac{1+r_se^{-2i\beta}}{1-r_se^{2i\beta}},\nonumber\\
\frac{t_sD_2}{M}&=&\frac{t_s}{1+r_se^{2i\beta}}.
\end{eqnarray}
Thus, defining
\begin{eqnarray}
\tilde{\cal K}\equiv\frac{t_s^2}{1+2r_s\cos{2\beta}+r_s^2}{\cal K},\label{eq:Ktilde}
\end{eqnarray}
we have
\begin{eqnarray}
S_h^\mathrm{BRSE}(f)=\frac{h_\mathrm{SQL}^2}{2}\left(\frac{1}{\tilde{\cal K}}+\tilde{\cal K}\right)\,.\label{eq:Sh2}
\end{eqnarray} 
The quantum noise level still cannot overcome the SQL. Equation~(\ref{eq:Ktilde}) tells us that the required power at the beamsplitter to achieve $\tilde{\cal K}=1$ is higher with the RSE technique. On the other hand, $\tilde{\cal K}$ can be rewritten as
\begin{eqnarray}
\tilde{\cal K}&=&\frac{t_s^2}{(1-r_s)^2+4r_s\cos^2{\beta}}\frac{8\omega_0I_0}{mL^2\Omega^2(\gamma^2+\Omega^2)}\nonumber\\
&=&\frac{1+r_s}{1-r_s}\frac{8\omega_0I_0}{mL^2\Omega^2(\tilde{\gamma}^2+\Omega^2)}\label{eq:Ktilde2}
\end{eqnarray}
with
\begin{eqnarray}
\tilde{\gamma}&=&\frac{1+r_s}{1-r_s}\gamma,
\end{eqnarray}
which means the bandwidth is expanded by $(1+r_s)/(1-r_s)$. Indeed, the RSE configuration is used with a lower $T$. With $T=0.004$ and $r_s=0.92$, the noise spectra shown in Fig~\ref{fig:SQL} can be realized using $\sim\!25$ times lower laser powers. This is an important advantage of the RSE configuration, especially if the laser power is limited by a thermal problem due to absorption in the input mirror substrate.

\section{Quantum noise reduction technique in KAGRA}\label{sec:3}

It is essential for KAGRA to try to overcome the SQL, as the sensitivity is mostly limited by quantum noise for its low mirror thermal noise with the cryogenic operation. As explained in Ref.~\cite{KAGRA}, KAGRA is designed to implement a couple of advanced techniques to reduce quantum noise and even overcome the SQL in the vicinity of a certain frequency. Let us introduce these techniques with practical challenges.

\subsection{Back-action evasion technique}

\begin{figure}[htbp]
	\begin{center}
		\includegraphics[width=8cm]{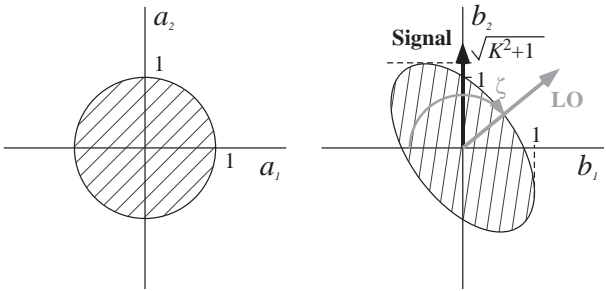}
	\caption{{\it Left}: Input coherent vacuum field. {\it Right}: Output squeezed vacuum field with a signal field in the phase quadrature and a local oscillator (LO) that works as a reference beam.}
	\label{fig:vacuum}
	\end{center}
\end{figure}

The first technique to reduce quantum noise is the back-action evasion (BAE) technique. Quantum noise has two components: (1) a fluctuation in the amplitude quadrature and (2) a fluctuation in the phase quadrature. There is no laser light at the anti-symmetric port so that only a vacuum field, the mean amplitude of which is zero, enters the interferometer. In the quadrature plane, the input vacuum field $a$ is represented by a circle, as the two components have equal fluctuation levels. On the other hand, the output field $b$ is a squeezed ellipse for the optomechanical coupling that converts an amplitude quadrature fluctuation into a phase amplitude fluctuation, as depicted in Fig.~\ref{fig:vacuum}. The BAE makes use of this squeezing to measure the signal in a quadrature where the noise ellipse is thin.

As will be explained in Sec.~\ref{sec:4}, a squeezed vacuum can be realized with a nonlinear crystal and a pump beam whose wavelength is exactly one half of the carrier light. The product of the upper sideband vacuum field and the pump beam forms the lower sideband vacuum field, and vice versa. A nonlinear crystal creates a correlation between the upper and lower sidebands, which are uncorrelated in the coherent vacuum. The output field of the interferometer is also squeezed because of the correlation between the upper and lower sidebands, but not in the same way. In the interferometer, a vacuum field in the amplitude quadrature couples with the carrier light to exert radiation pressure on the optics, which creates modulation sidebands in the phase quadrature. The quadrature fields consist of the upper and lower sideband fields, and therefore the optomechanical coupling creates a correlation between the upper and lower sidebands.

The BAE can be explained in a different way. With the standard measurement in the phase quadrature, the phase fluctuation is measured directly, and the amplitude fluctuation is measured via radiation pressure. With a measurement in a mixed quadrature, the amplitude fluctuation is measured in two ways: directly and via radiation pressure noise. Tuning the balance of the measurements, the amplitude fluctuation terms can be canceled. This can be regarded as a quantum feed-forward control.

\begin{figure}[htbp]
	\begin{center}
		\includegraphics[width=8cm]{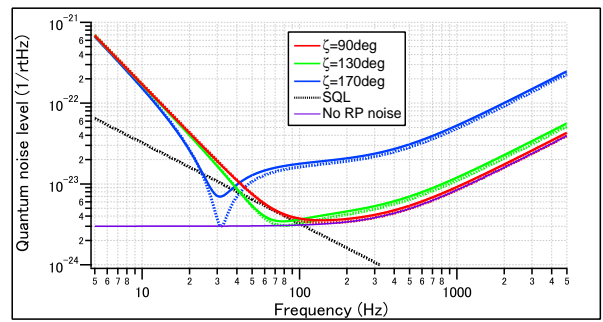}
	\caption{Quantum noise spectra of KAGRA in a tuned RSE configuration with three different readout phases. The dashed and solid curves show quantum noise with and without optical losses, respectively. With a non-phase-quadrature measurement, the noise spectrum can exceed the standard quantum limit (SQL), reaching the shot noise level at a certain frequency.}
	\label{fig:BAE}
	\end{center}
\end{figure}

Combining the outputs at $b_1$ and $b_2$ in the readout quadrature $\zeta$, we have the noise spectral density of a tuned RSE interferometer as follows:
\begin{eqnarray}
S_h^\mathrm{BAE}(f)=\frac{h_\mathrm{SQL}^2}{2}\frac{1+\left(\tilde{\cal K}+\cot{\zeta}\right)^2}{\tilde{\cal K}}\,.\label{eq:ShBAE}
\end{eqnarray} 
It corresponds to Eq.~(\ref{eq:Sh2}) with $\zeta=\pi/2$, i.e., in the phase-quadrature measurement. With the readout phase tuned to 
\begin{eqnarray}
\zeta=-\mathrm{Arccot}\tilde{\cal K}\,,\label{eq:cotK}
\end{eqnarray} 
the second term of the numerator of the right-hand side in Eq.~(\ref{eq:ShBAE}) disappears, as a result of which the quantum noise level reaches the shot noise level. In fact, the $\tilde{\cal K}$ term is a function of frequency, so that the sensitivity improvement is limited at around a certain frequency. Figure~\ref{fig:BAE} shows a few quantum noise spectra with different readout phases. The interferometer parameters are the ones used for KAGRA.

Although the signal-to-noise ratio is improved, the amount of the signal probed in the BAE technique is reduced by $\sin{\zeta}$. If there is an additional coherent (or squeezed in a different quadrature) vacuum field mixed with the output field, the signal-to-noise ratio is reduces more quickly than in the standard measurement. The dashed curves in Fig.~\ref{fig:BAE} show the noise spectra with the optical losses estimated in KAGRA.

There are two practical ways to provide a local oscillator (LO) and implement the BAE technique. One is a {\it DC readout}, and the other is a {\it balanced homodyne detection}. 

In the DC readout, the LO comes from the asymmetry of the main interferometer~\cite{DCreadout}. A loss imbalance of the test masses in two arm cavities creates an asymmetry in the cavity reflectivities, and a fraction of the light injected to the interferometer leaks to the anti-symmetric port. This component appears in the amplitude quadrature. A phase imbalance in two arm cavities, namely, an offset from the dark fringe operation, also allows a fraction of the beam leak through the anti-symmetric port. This component appears in the phase quadrature. The value of the amplitude quadrature component is basically fixed. It depends on the combination of the mirror maps in the two arm cavities. Our numerical simulation using 24 combinations of artificial mirror surface maps with the same power spectral density showed that the value of the amplitude quadrature component at the anti-symmetric port ranges from $60\,\mu$W to $16\,$mW~{\cite{Kumeta}\cite{YanoM}} with 515\,W at the beam splitter. In approximately 80\% of the cases, the amplitude quadrature component was larger than 1\,mW. If the amplitude component is lower than 1\,mW, it is hard to choose a DC readout phase far from $\pi/2$, as the total amount of the LO is so small that the requirements with respect to the scattering light, spatial higher-order modes, etc., would be too challenging. 

\begin{figure}[htbp]
	\begin{center}
		\includegraphics[width=6cm]{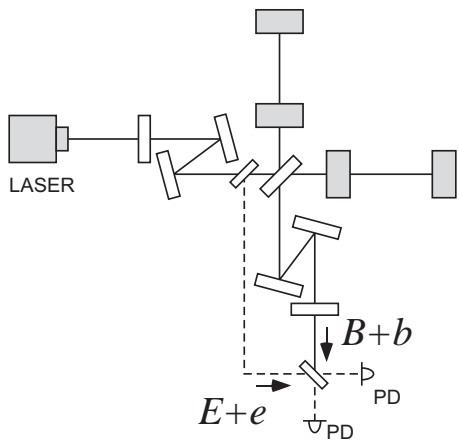}
	\caption{A balanced homodyne detection technique with a local oscillator beam taken out from the power recycling cavity.}
	\label{fig:BHD}
	\end{center}
\end{figure}

Although the DC readout does not always allow us to tune the readout phase, the balanced homodyne detection does. It requires a fraction of carrier light to leak out from the interferometer, preferably from the power-recycling cavity, which plays the role of a low-pass filter of laser noise, and an additional Mach-Zehnder interferometer to combine the light and the signal field at the anti-symmetric port (Fig.~\ref{fig:BHD}). The differential output of the two photodiodes is insensitive to the laser noise of the LO unless there is a DC component along with the signal field. Define the local oscillator field and its laser noise to be $E$ and $e$, respectively, and define a possible DC component and the output signal (+ vacuum) field at the anti-symmetric port to be $B$ and $b$, respectively. The differential output of the two photodiodes contains a signal term $E_1(b_1\cos{\zeta}+b_2\sin{\zeta})$ and a laser noise term $B_1(e_1\cos{\zeta}-e_2\sin{\zeta})+B_2(e_1\sin{\zeta}+e_2\cos{\zeta})$. To reduce the laser noise coupling, the LO should be low-pass-filtered with the power-recycling cavity, and the two arm cavities should be well balanced to reduce the $B$ component. A previous study in LIGO proposes to extract the LO from the anti-reflection coating side of the beam splitter~\cite{BHD}, which would be a good way to realize the balanced homodyne detection without introducing additional losses. They also propose to use another output mode-cleaner for the LO to further reduce laser noise.

In the case of KAGRA, we decided to try the DC readout scheme. It is less complicated than the balanced homodyne detector, and the 80\% figure represents a good chance. One challenge is that the requirement the output mode-cleaner has to satisfy becomes more severe, because the amount of the LO can be as low as $\sim\!1\,$mW. As a result, the finesse of the KAGRA's output mode-cleaner is twice as high as the LIGO's, and the round-trip length is also twice as long as the LIGO's.

For KAGRA, which has a carrier power of 780\,W at the beam splitter, the optimal readout phase to maximize the observation range of the neutron star binaries is $121^\circ$. The observation range (sky average) is 134\,Mpc with the phase measurement and 143\,Mpc with the optimal readout phase.
 
\subsection{Optical spring}

The second technique is a detuning of the SRC, which creates an optical spring that amplifies the signal in the vicinity of its resonant frequency. The key to realizing the optical spring is radiation pressure. A gravitational wave modulates the carrier light to generate a phase signal that arrives at the SRC and reenters the interferometer partially converted to an amplitude modulation to the carrier light. The carrier light couples to this amplitude modulation and exerts radiation pressure on the mirrors. The differential motion of the mirror due to the radiation pressure generates a phase signal. This optomechanical loop creates a virtual spring, which in the case of KAGRA is 3 km long. 

\begin{figure}[htbp]
	\begin{center}
		\includegraphics[width=8cm]{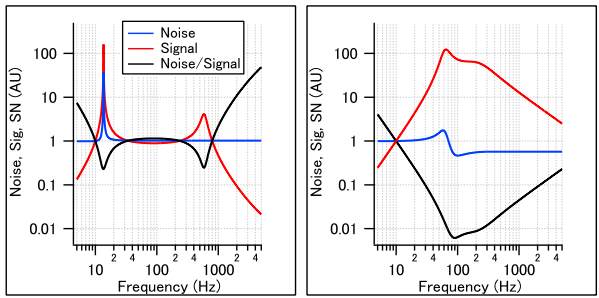}
	\caption{{\it Left}: Noise field amplitude, signal field amplitude, and the signal-to-noise ratio in a detuned RSE interferometer where the optical spring frequency is sufficiently far from the optical resonance. {\it Right}: Noise field amplitude, signal field amplitude, and the signal-to-noise ratio in KAGRA. Each curve is normalized by the value at 10\,Hz.}
	\label{fig:OS}
	\end{center}
\end{figure}

In Eq.~(\ref{eq:inputoutput}), the $1/M$ term represents the susceptibility of the detector to the external force on the test masses. With $\phi=\pi/2$, $M$ is always nonzero. With $\phi\neq0$ and $\beta,\Phi\ll1$, $M$ could be zero for a certain $\cal K$; in other words, at a certain frequency. At this frequency, namely, the optical spring resonance, both the signal field and the noise field are amplified. Looking closely, however, the noise field is amplified less by the optical spring, as a large fraction of the vacuum field entering from the anti-symmetric port is directly reflected by the signal-recycling mirror and does not experience the signal amplification. Thus, the signal-to-noise ratio is improved in the vicinity of this optical spring resonance. The left panel of Fig.~\ref{fig:OS} shows the signal response, the noise response, and the quantum noise spectrum of an interferometer (not KAGRA) in a detuned RSE configuration; the readout phase is set to the phase quadrature $\zeta=\phi$.

However, the above explanation does not hold when the optical resonance, the higher-frequency dip in the quantum noise spectrum, is not far from the optical spring frequency. The right panel of Fig.~\ref{fig:OS} shows the signal response, the noise response, and the quantum noise spectrum of KAGRA in a detuned RSE configuration; the detune phase is 3.5\,deg, and the readout phase is 131\,deg. The amplification is not as significant, but the noise field is squeezed in the vicinity of the optical spring frequency.

We can state that the quantum noise level is improved for the signal amplification with the optical spring in the former case, and for the squeezing in the vicinity of the optical spring frequency in the latter case. In either case, the quantum noise spectrum can exceed the SQL. 

Practical challenges to the realization of the optical spring are related to the control of the interferometer. First, an optical spring is an unstable spring with a negative imaginary part. Feedback control of the differential mode of the arm cavity lengths can easily solve this problem unless the optical spring frequency is higher than the frequency of some mechanical modes. Second, the SRC length needs to be set not to an integral multiple of the wavelength of the carrier light but to a slightly detuned one, so as to introduce an asymmetry in the response of the radio-frequency control fields around the carrier light. A preceding study~\cite{UedaCQG} has revealed that the oscillator phase noise would limit the KAGRA sensitivity below 500\,Hz, assuming a radio-frequency phase fluctuation of $-120\,\mathrm{dBc}/\sqrt{\mathrm{Hz}}$. One way to reduce the noise coupling is to inject a radio-frequency amplitude modulation to recover the symmetry between the upper and the lower sidebands. The oscillator phase noise level can be below the sensitivity curve with a safety factor of 10. Although there are several ways to produce an amplitude modulation, KAGRA has chosen to use the Mach-Zehnder interferometer with an electro-optic modulator in each arm; this is called a Mach-Zehnder modulation system~\cite{YamaKoh} and has been installed in the KAGRA input optics (see Fig.~\ref{fig:IFO}).

\subsection{Design sensitivity of KAGRA}\label{sec:3-3}

\begin{figure}[htbp]
	\begin{center}
		\includegraphics[width=8cm]{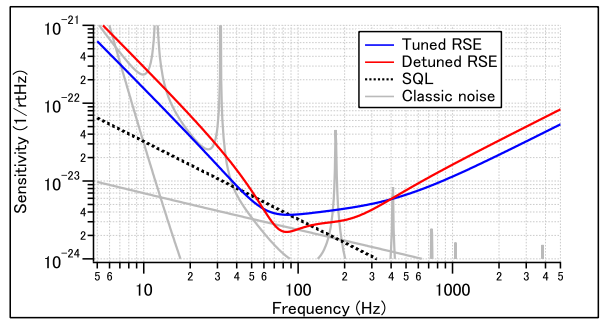}
	\caption{Quantum noise and classical noise curves in KAGRA.}
	\label{fig:sensitivity}
	\end{center}
\end{figure}

Figure~\ref{fig:sensitivity} shows the quantum noise spectra in the target sensitivity of KAGRA as of August 2017~\cite{KomoriJGW}. The detuned configuration is the primary goal, but a broadband configuration is also provided as an option for a slightly broader sensitivity. The arm cavity finesse and the signal-recycling mirror reflectivity have been determined to increase the observation range of a neutron star binary, the primary target source of KAGRA, and not to simultaneously narrow the observation bandwidth too much for other sources. This request resulted in the choice of high finesse and small detuning, while a moderate finesse and large detuning would have provided an even higher observation range.

\begin{figure}[htbp]
	\begin{center}
		\includegraphics[width=8cm]{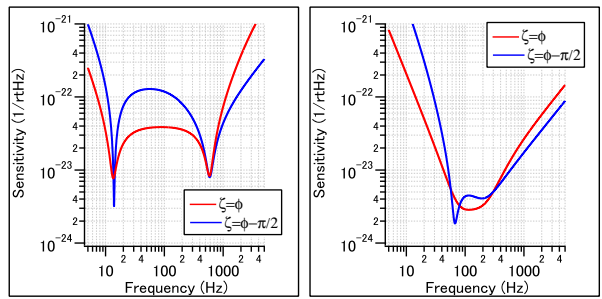}
	\caption{{\it Left}: Quantum noise spectrum of a detuned RSE interferometer with different readout phases. The parameters are the same as those shown in the left panel of Fig.~\ref{fig:OS}. {\it Right}: Quantum noise spectrum of a detuned RSE interferometer with different readout phases. The other parameters are the same as for KAGRA.}
	\label{fig:OS2}
	\end{center}
\end{figure}

The readout phase is selected to maximize the observation range for a neutron star binary. In the case of the detuned RSE, a non-phase-quadrature measurement does not provide a narrow band BAE sensitivity but reshapes the quantum noise spectrum. With $\zeta=\phi$, which is close to the phase-quadrature measurement, the low-frequency sensitivity is better, the high-frequency sensitivity is worse, and the optical spring dip is less steep than with $\zeta=\phi-\pi/2$ (see Fig.~\ref{fig:OS2}). 

The observation ranges for a neutron star binary with a tuned or a detuned configuration are shown in Table~\ref{tab:IR}. A gradual increase of the observation range can be seen. This reflects the fact that the sensitivity is still limited by quantum noise. KAGRA has devoted a lot of resources to cryogenics, and the thermal noise level will be lower than in the other telescopes of the same generation. It is important to implement the BAE and/or detuning to reduce quantum noise even if some challenges have to be overcome.
\begin{table}[htbp]
\begin{center}
\begin{tabular}{c|c}
Configuration&Observation range\\ \hline
Tuned RSE, $\zeta=\pi/2$&128\,Mpc\\ \hline
Tuned RSE, $\zeta$ is optimized&135\,Mpc\\ \hline
Detuned RSE, $\zeta=\phi$&140\,Mpc\\ \hline
Detuned RSE, $\zeta$ is optimized&153\,Mpc\\ \hline
\end{tabular}
\caption{Observation ranges for a neutron star binary with KAGRA in different configurations.}
\label{tab:IR}
\end{center}
\end{table}

\section{Advanced quantum noise reduction technique for KAGRA+}\label{sec:4}

We are aiming to reach the target sensitivity by 2022~\cite{LVRR}, and then expect to implement some even more advanced techniques to further improve the sensitivity. In this section, we introduce a few ideas proposed in the white paper for the KAGRA upgrade, which is called KAGRA+~\cite{WP}.

\subsection{Frequency dependent squeezing}\label{sec:FDSQ}

One of the most serious problems in KAGRA is the light mass of the sapphire mirrors. It is approximately one half of the mass of the silica mirrors used in LIGO or Virgo. The original plan was to use 50\,kg sapphire mirrors~\cite{LCGTdoc}, but it turned out to be impossible to produce such a large c-axis crystal bulk with a sufficient quality at the beginning of the project. The light mass leads to high quantum radiation pressure noise and hence to a high SQL. Frequency-dependent squeezing is a technique to reduce both shot noise and radiation pressure noise, and therefore it is equivalent to increasing both the mirror mass and the input laser power.

\begin{figure}[htbp]
	\begin{center}
		\includegraphics[width=6cm]{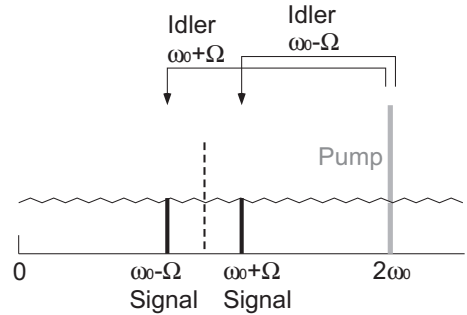}
	\caption{Through the nonlinear coupling with the pump beam at twice the frequency of the carrier light, the upper (lower) sideband signal field creates an idler field, the frequency of which coincides with that of the lower (upper) sideband signal field.}
	\label{fig:SQ}
	\end{center}
\end{figure}

Let us now examine a frequency-{\it dependent} squeezing. In Sec.~\ref{sec:2}, we introduced a coherent vacuum field entering from the anti-symmetric port $a$, the noise spectral density of which is 1 in either quadrature, as shown in Eq.~(\ref{eq:coherent}). The squeezing is a way of changing the equal balance of $S_{a1}$ and $S_{a2}$ while keeping the product $S_{a1}S_{a2}=1$. The key to increasing or decreasing the component is the correlation between the vacuum fields at the upper sideband and the lower sideband frequencies. The amplitude-quadrature and the phase-quadrature components are the common mode and differential mode fluctuations of those vacuum fields, and therefore the positive correlation decreases the phase-quadrature component, and the negative correlation decreases the amplitude-quadrature component. The uncertainty principle requires that the opposite quadrature component increase simulataneously so that the product of the two components remains unity in the absence of optical losses. A well-established way to realize the correlation is to use an optical parametric amplifier that consists of a nonlinear crystal and a pump beam at twice the frequency of the carrier light. With  the pump field $E_p(2\omega)$ and the signal field $E_s(\omega\pm\Omega)$ simultaneously injected to a crystal, the dielectric polarization density contains a nonlinear component $P(\omega\mp\Omega)$ that is proportional to $\epsilon_0\chi^{(2)}E_pE_s^*$. Here, $\chi^{(2)}$ is the second-order susceptibility. The output idler field $E_i(\omega\mp\Omega)$ proportional to this $P(\omega\mp\Omega)$ is then produced. The signal field frequencies coincide with the idler field frequency, and either the common mode or differential mode fluctuation can be suppressed while the other one is amplified (Fig.~\ref{fig:SQ}). 

A coherent vacuum field $(a_1,\,a_2)$ will be converted to a squeezed vacuum field $(a_1^\mathrm{SQ},\,a_2^\mathrm{SQ})$ with multiplication by a squeezing matrix ${\bf S}=((s,\,0),\,(0,\,1/s))$ and a rotation matrix: 
\begin{eqnarray}
\left (
\begin{array}{@{\,}c@{\,}}
a_1^\mathrm{SQ}\\
a_2^\mathrm{SQ}
\end{array}\!\right )=
\left (\!
\begin{array}{@{\,}cc@{\,}}
\cos{\theta}&-\sin{\theta}\\
\sin{\theta}&\cos{\theta}
\end{array}\!\right )\!\!
\left (\!
\begin{array}{@{\,}cc@{\,}}
s&0\\
0&1/s
\end{array}\!\right )\!\!
\left (
\begin{array}{@{\,}c@{\,}}
a_1\\
a_2
\end{array}\!\right )\,.
\end{eqnarray}

\begin{figure}[htbp]
	\begin{center}
		\includegraphics[width=8cm]{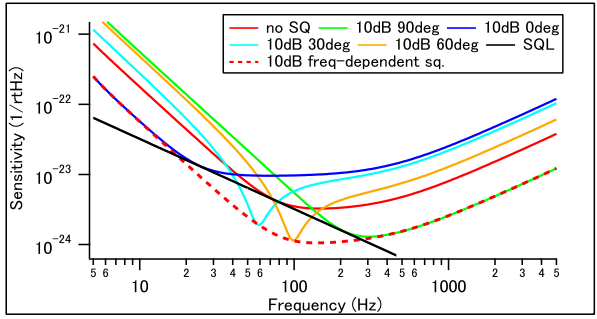}
	\caption{Quantum noise spectra with squeezing.}
	\label{fig:narrowSQ}
	\end{center}
\end{figure}

Let us fix the rotation angle $\theta$ to 0 and consider the lossless case. In this case, Eqs.~(\ref{eq:Sh})(\ref{eq:Sh2}) hold while the optomechanical coupling coefficient $\tilde{\cal K}$ changes as though the input power $I_0$ is increased by $s$:
\begin{eqnarray}
\tilde{\cal K}^\mathrm{SQ}&=&\frac{1+r_s}{1-r_s}\frac{8\omega_0I_0s}{mL^2\Omega^2(\tilde{\gamma}^2+\Omega^2)}\,.
\end{eqnarray}
If the squeezing rotation angle is $\pi/2$, the optomechanical coupling coefficient $\tilde{\cal K}$ changes as though the input power is decreased by $1/s$. The former case is called phase squeezing or simply squeezing, and the latter case is called amplitude squeezing or anti-squeezing. If the squeezing angle is chosen from between $0$ and $\pi/2$, the sensitivity curve will exceed the SQL in a narrow frequency band, as shown in Fig.~\ref{fig:narrowSQ}. 

There are many practical challenges to increasing the input laser power: thermal lensing of the mirrors~\cite{lensing}, tilt instability due to the radiation pressure force~\cite{tiltinstability}, parametric instability~\cite{PI}, etc. Phase squeezing is an attractive way to effectively increase the laser power. In fact, the technique has been tested and installed in advanced interferometers~{\cite{GEOSQ}\cite{LIGOSQ}}. The achieved squeezing level is limited by optical losses in the interferometer. The highest value so far has been recorded in GEO-HF, where a squeezing strength of $\sim\!6$\,dB was obtained, which is equivalent to increasing the input power by a factor of 4.

Now, let us consider frequency-dependent squeezing. If we can choose an appropriate squeezing angle corresponding to the best of all the curves shown in Fig.~\ref{fig:narrowSQ}, the sensitivity will be improved in a broad frequency band. A frequency-dependent rotation of the squeezing angle can be realized by injecting the squeezed vacuum into a detuned high-finesse optical resonator, namely a filter cavity~\cite{Kimble}.

The input-output relation of a detuned cavity without optomechanical coupling (no radiation pressure) is given by
\begin{eqnarray}
\left (
\begin{array}{@{\,}c@{\,}}
a_1^\mathrm{FC}\\
a_2^\mathrm{FC}
\end{array}\right )
=
\frac{1}{M_\mathrm{F}}
\left (
\begin{array}{@{\,}cc@{\,}}
F_{11}&F_{12}\\
F_{21}&F_{22}
\end{array}\right )
\left (
\begin{array}{@{\,}c@{\,}}
a_1^\mathrm{SQ}\\
a_2^\mathrm{SQ}
\end{array}\right )e^{2i\beta_\mathrm{F}}\ ,\label{eq:inout1}
\end{eqnarray}
with
\begin{eqnarray}
&&M_\mathrm{F}=1-2r_\mathrm{F}e^{2i\beta_\mathrm{F}}\cos{2\phi_\mathrm{F}}+r_\mathrm{F}^2e^{4i\beta_\mathrm{F}}\ ,\nonumber\\
&&F_{11}=F_{22}=-2r_\mathrm{F}\cos{2\beta_\mathrm{F}}+(1+r_\mathrm{F}^2)\cos{2\phi_\mathrm{F}}\ ,\nonumber\\
&&F_{12}=-F_{21}=-(1-r_\mathrm{F}^2)\sin{2\phi_\mathrm{F}}\ .\label{eq:FC}
\end{eqnarray}
Here $r_\mathrm{F}$, $\phi_\mathrm{F}$, and $\beta_\mathrm{F}=\Omega L_\mathrm{F}/c$ are the input mirror amplitude reflectivity, the detune phase, and the phase rotation of the squeezed field in a single path of the filter cavity, respectively. The length of the filter cavity is defined as $L_\mathrm{F}$.

Plugging in the output field of the filter cavity into the main interferometer, Eq.~(\ref{eq:inputoutput}), we obtain
\begin{eqnarray}
&&\left (
\begin{array}{@{\,}c@{\,}}
b_1\\
b_2
\end{array}\right )
=\frac{1}{M_\mathrm{F}}\left (
\begin{array}{@{\,}cc@{\,}}
B_{11}&B_{12}\\
B_{21}&B_{22}
\end{array}\right )
\left (
\begin{array}{@{\,}c@{\,}}
a_1^\mathrm{SQ}\\
a_2^\mathrm{SQ}
\end{array}\right )e^{2i(\beta+\beta_\mathrm{F})}\nonumber\\
&&\hspace{3cm}+\frac{\sqrt{2\tilde{\cal K}}}{h_\mathrm{SQL}}
\left (
\begin{array}{@{\,}c@{\,}}
0\\
h
\end{array}\right )e^{i\beta}\ ,\label{eq:inputoutputFC}
\end{eqnarray}
with 
\begin{eqnarray}
B_{11}&=&-2r_\mathrm{F}\cos{2\beta_\mathrm{F}}+(1+r_\mathrm{F}^2)\cos{2\phi_\mathrm{F}}\ ,\\
B_{12}&=&-(1-r_\mathrm{F}^2)\sin{2\phi_\mathrm{F}}\ ,\\
B_{21}&=&-\tilde{\cal K}\{-2r_\mathrm{F}\cos{2\beta_\mathrm{F}}+(1+r_\mathrm{F}^2)\cos{2\phi_\mathrm{F}}\}\nonumber\\
&&\ \ \ \ -(1-r_\mathrm{F}^2)\sin{2\phi_\mathrm{F}}\ ,\label{eq:B21}\\
B_{22}&=&-2r_\mathrm{F}\cos{2\beta_\mathrm{F}}+(1+r_\mathrm{F}^2)\cos{2\phi_\mathrm{F}}\nonumber\\
&&\ \ \ \ +\tilde{\cal K}(1-r_\mathrm{F}^2)\sin{2\phi_\mathrm{F}}\ .
\end{eqnarray}
If $B_{21}$ can be made zero, quantum noise can be reduced as much as we want by phase-squeezing the input field. Let us assume that radiation pressure noise is low at frequencies higher than the cavity pole. We can employ the following approximation:
\begin{eqnarray}
\tilde{\cal K}&=&\frac{t_s^2}{1+2r_s\cos{2\beta}+r_s^2}\frac{2\tilde{\gamma}^4}{\Omega^2(\tilde{\gamma}^2+\Omega^2)}\frac{4\omega_0I_0}{mL^2\gamma^4}\nonumber\\
&\simeq&\frac{1-r_s}{1+r_s}\frac{8\omega_0I_0}{mL^2\gamma^4}\frac{\tilde{\gamma}^2}{\Omega^2}\ .
\end{eqnarray}
At low frequencies, the following approximation holds: $\cos{2\beta_\mathrm{F}}\simeq1-2\Omega^2L_\mathrm{F}^2/c^2$. The right-hand side of Eq.~(\ref{eq:B21}) has a term that does not depend on $\Omega$ and a term that is proportional to $1/\Omega^2$, both of which should be zero to reduce quantum noise at all the frequencies. Because the latter term is zero, $2r_\mathrm{F}=(1+r_\mathrm{F}^2)\cos{2\phi_\mathrm{F}}$. That is,
\begin{eqnarray}
\cos{2\phi_\mathrm{F}}=\frac{2r_\mathrm{F}}{1+r_\mathrm{F}^2}\ , \ \ \ \sin{2\phi_\mathrm{F}}=\frac{1-r_\mathrm{F}^2}{1+r_\mathrm{F}^2}\ .\label{eq:phase}
\end{eqnarray}
The condition to make $B_{21}$ zero is then $2r_\mathrm{F}\tilde{\cal K}\cos{2\beta_\mathrm{F}}=2r_\mathrm{F}\tilde{\cal K}-(1-r_\mathrm{F}^2)^2/(1+r_\mathrm{F}^2)$, which leads to
\begin{eqnarray}
\gamma_F\equiv\frac{T_\mathrm{F}c}{4L_\mathrm{F}}\simeq\frac{\Omega_{rp}}{\sqrt{2}}\ .\label{eq:fcp}
\end{eqnarray}
Here, $\gamma_\mathrm{F}$ is the cavity pole of the filter cavity, and $\Omega_\mathrm{rp}$ is the touching frequency of radiation pressure noise to the SQL, which is given as follows if $\Omega_\mathrm{rp}\ll\tilde{\gamma}$:
\begin{eqnarray}
\Omega_\mathrm{rp}&=&\sqrt{\frac{1-r_s}{1+r_s}\frac{8\omega I_0}{mL^2\gamma^2}}\ .
\end{eqnarray}
Equation~(\ref{eq:fcp}) is the essential equation for the design of the filter cavity. If the radiation pressure noise curve touches the SQL at 100\,Hz without squeezing and the filter cavity length is 30\,m, the finesse of the filter cavity must be 18,000. Stable control of such a high-finesse cavity is a challenge to the realization of the frequency-dependent squeezing. A demonstration experiment for the KAGRA filter cavity has been conducted at NAOJ~\cite{NAOJFC}. 

With 10-dB frequency-dependent squeezing and no optical losses, the effective test mass increases by a factor of $\sqrt{10}$, and the effective intra-cavity power increases by a factor of 10. The low mass and the low intra-cavity power are two big problems for the cryogenic interferometer, making frequency-dependent squeezing an attractive option for the KAGRA+.

An alternative method for frequency-dependent squeezing has been recently proposed by Ma {\it et al.}~\cite{Yiqiu} This method proposes to use an existing arm cavity as a filter cavity to lower the cost of building a new vacuum system. The key to realizing frequency-dependent squeezing is the use of an entangled squeeze field to be injected to an arm cavity. It has been considered as an alternative option for KAGRA+.

\subsection{Long SRC}

\begin{figure}[t]
	\begin{center}
		\includegraphics[width=8cm]{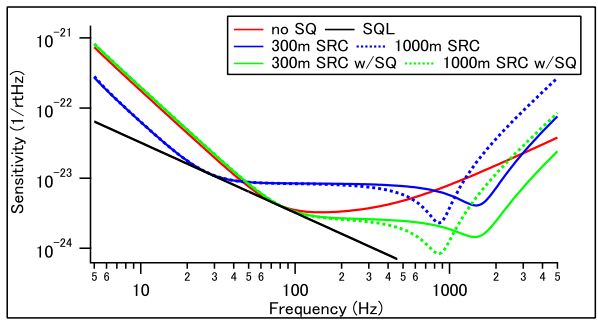}
	\caption{Quantum noise spectra with a long signal-recycling cavity (SRC). The dashed curves are with 10-dB frequency-independent squeezing.}
	\label{fig:LSRC}
	\end{center}
\end{figure}

Up to now the phase rotation gained by the signal sideband field, $\Phi$, has been ignored. The influence of this phase rotation starts to appear in the observation band when the length and/or the finesse of the SRC length increases~\cite{longSRC}. Figure~\ref{fig:LSRC} shows a few example spectra with different SRC lengths. The signal-recycling mirror reflectivity has been increased from 85\% to 98\% so that the influence of the phase rotation is more significant.

Here, we plot sensitivity curves with 10-dB frequency-{\it independent} squeezing. It is worth noting that the squeeze angle does not rotate around the dip frequency unlike the detuned SRC, so that frequency-dependent squeezing is not required to improve the sensitivity at high frequencies. A filter cavity is indeed useful to improve the sensitivity at low frequencies, just as we showed in Sec.~\ref{sec:FDSQ}.

One challenge to the realization of those sensitivity curves would be the extension of SRC in the tunnel. The current SRC length of KAGRA is 66.6\,m with two folding mirrors inside. It is not easy to further increase the length in the current vacuum system.

\begin{figure}[t]
	\begin{center}
		\includegraphics[width=8cm]{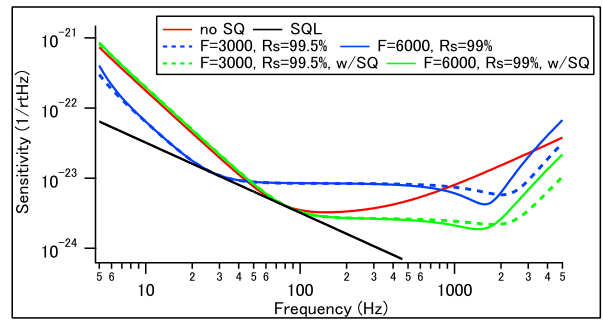}
	\caption{Quantum noise spectra with higher finesse arm cavities to emphasize the long signal-recycling cavity (SRC) effect. The dashed curves represent 10-dB frequency-independent squeezing.}
	\label{fig:LSRC2}
	\end{center}
\end{figure}

In fact, the same behavior can be observed with the current SRC length if we increase the finesse of the arm cavities. Figure~\ref{fig:LSRC2} shows example sensitivity curves with the 66.6\,m SRC. Here, we change the power-recycling gain to approximately tune the intra-cavity power and change the signal-recycling gain to approximately tune the shot noise floor level. The sensitivity with $F=3000$ almost coincides with the sensitivity curve with $L=300$\,m in Fig.~\ref{fig:LSRC} in the case without squeezing. The parameters used to obtain the spectra are summarized in Table~\ref{tab:LSRC2}. The round-trip loss of 100\,ppm in each arm cavity is included but no other optical losses are included.

\begin{table}[htbp]
\begin{center}
\begin{tabular}{c|c|c|c}
&finesse&PRM reflectivity&SRM reflectivity\\ \hline
KAGRA&1500&90\%&85\%\\ \hline
example 1&3000&80\%&99.5\%\\ \hline
example 2&6000&60\%&99\%\\ \hline
\end{tabular}
\caption{Parameters used in the calculation for Fig.~\ref{fig:LSRC2}.}
\label{tab:LSRC2}
\end{center}
\end{table}

\section{Summary}

In this article, we summarize the quantum noise reduction techniques that have been implemented in KAGRA and the promising quantum noise reduction techniques proposed for KAGRA+. Because the sensitivity of KAGRA is mostly limited by quantum noise, thanks to the cryogenic operation, quantum noise reduction is essential to improve the sensitivity of the gravitational-wave observation. For the current KAGRA, the BAE technique and the optical spring are to be implemented. We encountered some practical difficulties but have found a solution to each of them. For the future upgrade, frequency-dependent squeezing and the long SRC with frequency-independent squeezing would be the most promising candidates. Further investigation for the actual implementation would be needed for the actual implementation.

\section{Acknowledgement}

This work was supported by MEXT, JSPS Leading-edge Research Infrastructure Program, JSPS Grant-in-Aid for Specially Promoted Research 26000005, JSPS Core-to-Core Program A. Advanced Research Networks, JSPS KAKENHI grant number JP17H02886, the Mitsubishi Foundation, and the JST CREST grant number JPMJCR1873. We would like to thank Editage [http://www.editage.com] for editing and reviewing this manuscript for English language.

\bibliographystyle{junsrt}

\begin{thebibliography}{99}
\thispagestyle{headings}
\bibitem{KAGRA} K.~Somiya for KAGRA Collaboration, Class.~Quantum~Grav., \textbf{29}, 124007 (2012).
\bibitem{BAE} W.~G.~Unruh, p.~647 in {\it Quantum Optics, Experimental Gravitation, and Measurement Theory,} edited by P.~Meystre and M.~O.~Scully, Plenum Press, New York (1982)
\bibitem{OpticalSpring} A.~Buonanno and Y.~Chen, Phys.~Rev.~D, \textbf{64}, 042006 (2001)
\bibitem{Kimble} H.~J.~Kimble, Y.~Levin, A.~B.~Matsko, K.~S.~Thorne, and S.~P.~Vyatchanin, Phys.~Rev.~D, \textbf{65}, 022002 (2001)
\bibitem{SQL} V.~B.~Braginsky and F.~Ya.~Khalili, Rev.~Mod.~Phys. \textbf{68}, 1 (1996)
\bibitem{Mizuno} J.~Mizuno, K.~A.~Strain, P.~G.~Nelson, J.~M.~Chen, R.~Schilling, A.~R\"{u}diger, W.~Winkler, and K.~Danzmann, Phys.~Lett.~A \textbf{175}, 273 (1993)
\bibitem{longSRC} H.~Miao, H.~Yang, R.~X.~Adhikari, and Y.~Chen, Class.~Quantum~Grav., \textbf{31}, 165010 (2014)
\bibitem{WP} KAGRA Future Planning Committee, ``KAGRA FPC White Paper,'' JGW-M1909590-v11 (2019)
\bibitem{DCreadout} K.~Somiya, Y.~Chen. S.~Kawamura, and N.~Mio, Phys.~Rev.~D, \textbf{73}, 122005 (2006)
\bibitem{Kumeta} A.~Kumeta, C.~Bond, and K.~Somiya, Opt.~Rev., \textbf{22}, 149-152 (2015)
\bibitem{YanoM} K.~Yano, ``Design study and prototype experiment for the KAGRA output mode-cleaner,'' master's thesis, Tokyo Institute of Technology (2016)
\bibitem{BHD} P.~Fritschel, M.~Evans, and V.~Frolov, Opt.~Exp., \textbf{22}, 4224-4234 (2014)
\bibitem{UedaCQG} S.~Ueda, N.~Saito, D.~Friedrich, Y.~Aso, and K.~Somiya, Class.~Quantum~Grav., \textbf{31}, 095003 (2014)
\bibitem{YamaKoh} K.~Yamamoto {\it et al.}, Class.~Quantum~Grav., accepted (2019)
%\bibitem{YamaKoh} K.~Yamamoto {\it et al.}, Class.~Quantum~Grav., \textbf{??}, 000000 (2019)
\bibitem{KomoriJGW} ``Latest estimated sensitivity of KAGRA (v201708),'' JGW-T1707038-v9 (2017)
%; https://gwdoc.icrr.u-tokyo.ac.jp/cgi-bin/private/DocDB/ShowDocument?docid=7038
\bibitem{LVRR} B.~P.~Abbott {\it et al.} (VIRGO, KAGRA, LIGO Scientific), Living~Rev.~Rel., \textbf{21}, 3 (2018)
\bibitem{LCGTdoc} LCGT Collaboration, ``LCGT Design Document ver.3,'' JGW-T0400030-v4 (2009)
%; https://gwdoc.icrr.u-tokyo.ac.jp/cgi-bin/private/DocDB/ShowDocument?docid=30
\bibitem{lensing} P.~Willems and for the LIGO Scientific Collaboration, Frontiers in Optics 2009/Laser Science XXV/Fall 2009 OSA Optics \& Photonics Technical Digest, AOThA5 (2019)
\bibitem{tiltinstability} J.~A.~Sidles and D.~Sigg, Phys.~Lett.~A \textbf{354}, 167 (2006)
\bibitem{PI} V.~B.~Braginsky, S.~E.~Strigin, and S.~P.~Vyatchanin, Phys.~Lett.~A \textbf{287}, 331 (2001)
\bibitem{GEOSQ} J.~Abadie {\it et al.}, Nature~Phys., \textbf{7}, 962 (2011)
\bibitem{LIGOSQ} J.~Aasi {\it et al.}, Nature~Photonics, \textbf{7}, 613 (2013)
%\bibitem{VirgoSQ}
\bibitem{NAOJFC} E.~Capocasa {\it et al.}, Phys.~Rev.~D, \textbf{98}, 022010 (2018)
\bibitem{Yiqiu} Y.~Ma {\it et al.}, Nature~Phys., \textbf{13}, 776-780 (2017)
\end{thebibliography}
\pagestyle{headings}

\end{document}